\documentclass[epj]{svjour}
\usepackage{graphics}
\usepackage{autobreak}
\usepackage[T1]{fontenc}
\usepackage{mathptmx}      
\usepackage{flushend}
\usepackage{dcolumn}
\usepackage[numbers,sort&compress]{natbib}
\usepackage[colorlinks,citecolor=blue,urlcolor=blue,linkcolor=blue]{hyperref}
\begin{document}
\title{The effect of gluon condensate on imaginary potential and thermal width from holography}
\author{Yan-Qing Zhao\inst{1,}\thanks{email: yanqzhao@qq.com} \thanks{The first two authors contribute equally} \and Zhou-Run Zhu\inst{2,}\thanks{email: zhuzhourun@mails.ccnu.edu.cn} \and Xun Chen\inst{2,}\thanks{email: chenxunhep@qq.com}
}                     

\institute{College of Science, China Three Gorges University, Yichang 443002, China, \and Institute of Particle Physics and Key Laboratory of Quark and Lepton Physics (MOS), Central China Normal University, Wuhan 430079, China}
\date{Received: date / Revised version: date}
\abstract{
By the use of the gauge/gravity duality, we calculate the imaginary part of heavy quarkonium potential and thermal width with the effect of gluon condensate which is absent in AdS$_{5}$ background. Our results show that the dropping gluon condensate reduces the absolute value of imaginary potential and therefore decreases the thermal width both in "exact" and "approximate" approach implying that the heavy quarkonium has a weaker bound with the increase of gluon condensate. In addition, the thermal width will disappear at a critical condensate value, which indicates the dissociation of quarkonium. We conclude that increasing gluon condensate will lead to easier dissociation of heavy quarkonium for fixed temperature.
} 
\maketitle

\section{Introduction}
\label{intro}
\setlength{\parskip}{1em}
It is well-known that the experiments, RHIC and LHC, have found a new state of matter which is called as quark gluon plasma(QGP) produced by the heavy ion collisions \cite{Adams:2005dq,Adcox:2004mh}. Heavy quark-antiquark pair can be regarded as one of probes in the process of quark-gluon plasma(QGP) formation because the dissolution of quarkonium implies the occurrence of deconfinement phase transition\cite{Shuryak:1980tp,Matsui:1986dk}. We usually use heavy quarkonium potential $V_{Q\bar{Q}}$ to describe the interaction energy between quark and anti-quark. It is found that the potential may be in possession of a imaginary part at non-zero temperature, which is closely related to the decouple of heavy quarkonium, and it is believed that the dissolution of quarkonium is not because the binding energy disappear but because the reduced binding energy becomes as big as the thermal width which can be calculated by imaginary potential\cite{Laine:2006ns,Brambilla:2010vq,Noronha:2009da,Fadafan:2013bva,Finazzo:2013aoa,Fadafan:2013coa,Braga:2016oem,Thakur:2016cki}. So far, there are two main mechanisms for quarkonium dissociation or the appearance of imaginary potential, one is from the Landau damping of approximately static fields\cite{Laine:2006ns,Laine:2007gj,Beraudo:2007ky} and the other is its color singlet to color octet thermal break up\cite{Brambilla:2008cx}.

In the past few years, a lot of works about the imaginary part of the heavy quarkonium potential have been done in a weakly coupled theory\cite{Dumitru:2009fy,Margotta:2011ta,Chandra:2010xg,Escobedo:2013tca}. However, QCD theory is a strong coupling theory. Gauge/gravity duality\cite{Shuryak:2004cy,Maldacena:1997re,Gubser:1998bc,Witten:1998qj}, breaking the conformal symmetry at low energy, provides a very important tool to research the properties of hadron physics in a strong coupling systems\cite{Erlich:2005qh,deTeramond:2005su,DaRold:2005mxj,Babington:2003vm,Kruczenski:2003uq,Sakai:2004cn,Sakai:2005yt,Csaki:2006ji,
 Huang:2007fv,Gherghetta:2009ac,Kelley:2010mu,Sui:2009xe,Sui:2010ay,Li:2012ay,Li:2013oda,Chen:2015zhh,Evans:2006ea,Xiong:2019wik,Fang:2015ytf,Xie:2019soz}. Many scholars used holographic approach to study imaginary potential by considering various background like taking into account of chemical potential\cite{Zhang:2016tem}, they observed that the presence of the chemical potential decreases the dissociation length. In addition, many researches show that the process of heavy ion collisions will produce strong magnetic field\cite{Zhong:2014cda,Skokov:2009qp,Gusynin:1994re,Bali:2011qj,Gusynin:1995nb,Bali:2012zg,Voronyuk:2011jd,DElia:2010abb,Shushpanov:1997sf,Kharzeev:2012ph,Tuchin:2013ie,Bruckmann:2013oba,Baym:1995fk,Guo:2019joy,Chen:2019qoe}, so a numbers of papers based on this have been published, such as Ref.\cite{Zhang:2018fpe}, which shows that increasing magnetic field enhances the imaginary potential and decreases thermal width. In Ref.\cite{Zhu:2019ujc}, the energy loss of heavy and light quarks has been calculated in holographic model. Moving case was also taken into account in\cite{Chen:2017lsf,Finazzo:2014rca} which drew a conclusion that slowly moving quarkonium are less stable than static case.

When we consider a full quantum theory of QCD, a non-zero the trace of the energy-momentum tensor is manifested, since there is an anomaly, implying a nonzero gluon condensate. One can calculate the gluon condensate by trace anomaly\cite{Colangelo:2013ila,Leutwyler:1992cd,Castorina:2007qv} as follows
 \begin{equation}\label{eq:xx00}
   \Delta G_2(T)=G_2(T)-G_2(0)=-(\epsilon(T)-3P(T)),
\end{equation}
where $G_2(T)$ denotes the gluon condensate of limited temperature, $G_2(0)$, being equal to the condensate value at deconfinement transition temperature, is the condensate vale of zero temperature, $\epsilon(T)$ is the energy density, $p(T)$ is the pressure of QGP system. However, recent lattice calculations based on a QCD sum rule method have shown that the gluon condensate behaves a rapid change around $T_c$\cite{Morita:2007hv,Boyd:1996ex}. If we ignore shift in the width of the quarknia, this change will result in a reducing heavy quarkonium mass around $T_c$. Therefore, one can find that the gluon condensate is quite sensitive to the deconfinement phase transition of QCD and could be regarded as the signal to study phase transition. A lot of works on gluon dynamics have been investigated in holographic model\cite{Csaki:2006ji,Kim:2007qk,Kim:2008ax,Zhang:2019cxu,Ali-Akbari:2014vpa,Li:2013oda,Chen:2015zhh,Gursoy:2008bu,Kopnin:2011si,ChenXun:2019zjc,Chen:2019rez}. Inspired by this, we research the impact of gluon condensate on imaginary potential and thermal width in this work.

 The heavy quark anti-quark potential, at finite temperature, can be drawn from the vacuum expectation value in Wilson loop operator\cite{Wilson:1974sk,Gervais:1979fv,Polyakov:1980ca}. From the standpoint of AdS/CFT correspondence, the value of $\langle W(C)\rangle$ in the limit of large $N_c$ of strongly coupled 4-dimensional gauge theory is dual to gravity in the 5-dimensional bulk geometry. So we can get the stringy partition function. The stringy partition function in the classical gravity approximation is approximately equal to the exponent of the imaginary number unit $i$ multiplied by the classical string action, which could be regarded as the Nambu-Goto action. Then, the imaginary part of potential and thermal width can be calculated.

In this paper, we will take saddle point approximation approach to study the effect of the gluon condensate on imaginary part of heavy quark-antiquark potential, which leads to some restrictive conditions. Besides, the calculating results show that there is a critical condensate value, where the thermal width will disappear implying the dissociation of quarkonium. The rest of this paper is organized as follows. In section~\ref{sec:holo}, we briefly review the holographic model with the effect of gluon condensate. In section~\ref{sec:GC}, the results of the impact of gluon condensate on imaginary potential and thermal width are displayed. In section~\ref{sec:Con}, conclusion and further discussion are given.

\section{The holographic model}
\label{sec:holo}
To begin with, we briefly review the 5-dimensional gravity background in Minkowski space with a dilaton~\cite{Nojiri:1998yx}
\begin{equation}\label{eq:x1}
S=\frac{1}{2k^2} \int d^5x\sqrt{g} \bigg(\mathcal{R}+\frac{12}{L^2}-\frac{1}{2}\partial _\mu \phi \partial ^\mu \phi\bigg),
\end{equation}
where $k$ is the 5-dimensional gravitational coupling, $L$ is the radius of the asymptotic AdS$_5$ spacetime. $\phi$ is a massless scalar coupled with the gluon operator $G_{\mu\nu}G^{\mu\nu}$. By solving the above action, we can get dilaton equation of motion(EOM) and the Einstein equation. The solutions can be solved from EOM with a suitable metric ansatz. One is the dilaton-wall solution\cite{Nojiri:1998yx,Kehagias:1999tr},

\begin{align}\label{eq:x2}
  ds^2&=\frac{L^2}{z^2}(\sqrt{1-c^2z^8}(d\vec{x}^2+dt^2)+dz^2),
  \\
  \phi(z)&=\sqrt{\frac{3}{2}}log\bigg(\frac{1+cz^4}{1-cz^4}\bigg)+\phi_0,
\end{align}
where $\phi_0$ is a constant and $\vec{x}=x_1,x_2,x_3$ are orthogonal spatial boundary coordinates. $z$ denotes the 5th dimension radial coordinate. We assume the asymptotically AdS$_5$ boundary for gravity dual is at $z=0$. On top of this, $c=1/z_0^4$. The value of $1/z_0$ can be determined by the mass of the lowest meson\cite{Kim:2007rt} or lightest glueball\cite{Kehagias:1999tr}.

 The other metric is the dilaton black hole solution~\cite{Babington:2003vm,Kruczenski:2003be}.
\begin{equation}\label{eq:x3}
 ds^2=\frac{L^2}{z^2}\bigg(A(z)d\vec{x}^2-B(z)dt^2+dz^2\bigg),
\end{equation}
where
\begin{align}\label{eq:x4}
  A(z)&=(1+fz^4)^{(f+a)/2f}(1-fz^4)^{(f-a)/2f},  \\
  B(z)&=(1+fz^4)^{(f-3a)/2f}(1-fz^4)^{(f+3a)/2f}, \\
  f^2&=a^2+c^2,
\end{align}
and the corresponding dilaton profile is
\begin{equation}\label{eq:x5}
  \phi(z)=\frac{c}{f}\sqrt{\frac{3}{2}}log\bigg(\frac{1+fz^4}{1-fz^4}\bigg)+\phi_0.
\end{equation}

A Taylor expansion is taken near the boundary $z=0$ for the two dilatons above, one can get
\begin{equation}\label{eq:1}
  \phi(z)=\phi_0+\sqrt{6}cz^4+\cdots.
\end{equation}
we expect the constant piece to correspond to the source for the operator $TrG^2$ and coefficient of the $z^4$ to give the gluon condensate, see Ref.\cite{Csaki:2006ji} for a detailed discussion.

 Note that the position of the singularity is determined by $f=z_c ^{-4}$, where $z_c$ is an IR cut-off. One can easily find that the dilaton black hole solution becomes the AdS black hole solution when $c=0$, and the solution reduces to the dilaton wall solution with $a=0$. In addition, a Hawking-Page transition  occurs between the dilaton wall background and the dilaton black hole background at some critical value of $a$. Therefore, the dilaton wall solution corresponds to confined phase, and the dilaton black hole solution describes deconfined phase. More details can be seen from\cite{Kim:2007qk}.
 Next, we convert $z$ coordinate to $r$ coordinate by taking $r=\frac{L^2}{z}$, then the metric~\eqref{eq:x3} is
\begin{equation}\label{eq:x6}
  ds^2=\frac{r^2}{L^2}\bigg(A(r)d\vec{x}^2-B(r)dt^2\bigg)+\frac{L^2}{r^2}dr^2,
\end{equation}
with
\begin{equation}\label{eq:x7}
 \phi(r)=\frac{c}{f}\sqrt{\frac{3}{2}}log\bigg(\frac{1+fr^{-4}}{1-fr^{-4}}\bigg)+\phi_0,
\end{equation}
where
\begin{align}\label{eq:x8}
  A(r)&=(1+fr^{-4})^{(f+a)/2f}(1-fr^{-4})^{(f-a)/2f},  \\
  B(r)&=(1+fr^{-4})^{(f-3a)/2f}(1-fr^{-4})^{(f+3a)/2f}, \\
  f^2&=a^2+c^2.
\end{align}

Parameter $a$ is related to the temperature, $a=(\pi T^4)/4$. The location of horizon $r_h$ = $a^\frac{1}{4}$. Note that the presence of the cut-off, the range becomes $r_{f} < r < \infty$. In addition, we set the value of gluon condensate $0\le c\le 0.9 GeV^4$\cite{Kim:2008ax,Zhang:2019cxu}. We set $L=1$ in this paper.

\section{The effect of gluon condensate on imaginary potential}
\label{sec:GC}
In this section, following the step in~\cite{Fadafan:2013bva,Finazzo:2013aoa,Giataganas:2013lga}, we will calculate the imaginary potential and thermal width under the background metric~\eqref{eq:x6}. Now, we consider a rectangular Wilson loop, whose length of short side is $L$ in the spatial direction and a long side is $\mathcal{T}$ along a time direction. Set a dipole located in the direction of short side and the quarks are located at $x_1=\pm \frac{L}{2}$. Then, we use the string worldsheet coordinates
\begin{equation}\label{eq:x9}
  t=\tau,\quad  x_1=\sigma, \quad  x_2=x_3=const,\quad  r=r(\sigma).
\end{equation}

The Nambu-Goto action of the string is
\begin{equation}\label{eq:x10}
 S_{NG}=-\frac{1}{2\pi \alpha'}\int d\sigma\,d\tau \sqrt{-g},
\end{equation}
where $\alpha'$ is related to 't Hooft coupling as $\frac{1}{\alpha'}=\sqrt{\lambda}$. $g$ is the determinant of the induced metric given by
\begin{equation}\label{eq:x12}
  g_{00}=r^2B(r),\quad  g_{01}=g_{10}=0,\quad    g_{11}=r^2A{(r)}+\frac{r'^2}{r^2}.
\end{equation}

 In the calculation of imaginary potential, we will not consider the impact of dilaton. Then we can get the lagrangian density
\begin{equation}\label{eq:x13}
 \mathcal{L}=\sqrt{U(r)+V(r)r'^2},
\end{equation}
where
\begin{equation}\label{eq:x14}
  U(r)=r^4A(r)B(r),\quad  V(r)=B(r),
\end{equation}
and the prime represents derivative with respect to $\sigma$. Since the Lagrangian in this paper has the same form as that in Refs.\cite{Fadafan:2013bva,Finazzo:2013aoa,Giataganas:2013lga}, we will straightforwardly give the formulas of separation distance, imaginary potential and thermal width according to Refs.\cite{Fadafan:2013bva,Finazzo:2013aoa,Giataganas:2013lga}, as follows
\begin{equation}\label{eq:x18}
  L=2{\int_{r_0}}^\infty dr \sqrt{\frac{U(r_0)\,V(r)}{U(r)\,(U(r)-U(r_0))}},
\end{equation}
\begin{align}\label{eq:x30}
\begin{autobreak}
  ImV_{Q\bar{Q}}
  =-\frac{1}{2\sqrt{2} \alpha'}
  \,\sqrt{V(r_*)}
  \,\bigg(\frac{U'(r_*)}{2U''(r_*)}
  -\frac{U(r_*)}{U'(r_*)}\bigg),
\end{autobreak}
\end{align}
\begin{equation}\label{eq:x33}
  \frac{\Gamma_{Q\bar{Q}}}{T}=-\frac{4}{(a_0\,T)^3}\int d\omega\,{\omega^2}\,e^{\frac{-2\omega}{a_0\,T}}\,\frac{ImV_{Q\bar{Q}}(\omega)}{T},
\end{equation}
where $\omega=LT$. A upper and lower bound need be given for this integral. A upper limit can be easily obtained by the maximum of $LT$, while a lower bound can be derived by this condition $\sigma\ge0$ \cite{Fadafan:2013bva,Finazzo:2013aoa,Giataganas:2013lga}. So the range of this integral is $LT_{min}\le\omega\le LT_{max}$.
\par
\begin{figure}
\resizebox{0.5\textwidth}{!}{%
  \includegraphics{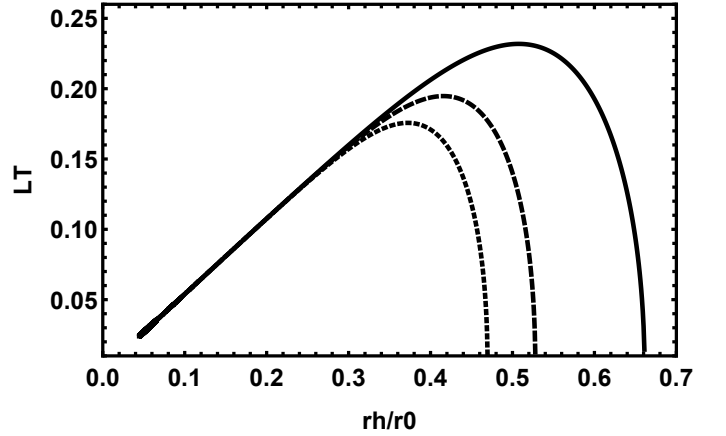}
}
\caption{$LT$ as a function of $rh/r_0$ for some choices of $c$ with a fixed temperature $T=0.2GeV$. Solid line represents $c=0.2GeV^4$, dashed line denotes $c=0.5GeV^4$, dotted line is $c=0.8GeV^4$.}
\label{fig:1}       
\end{figure}

 With all the preparations in place, $LT$ versus $rh/r_0$ for different values of $c$ is shown in Fig.~\ref{fig:1}. One can find clearly that increasing $c$ leads to decreasing $LT_{max}$. Here, we define $LT_{max}$ as the dissolution length of heavy quarkonium. Namely, the increase of gluon condensate leads to a easier dissociation of heavy quarkonium.

\begin{figure}
\resizebox{0.5\textwidth}{!}{%
  \includegraphics{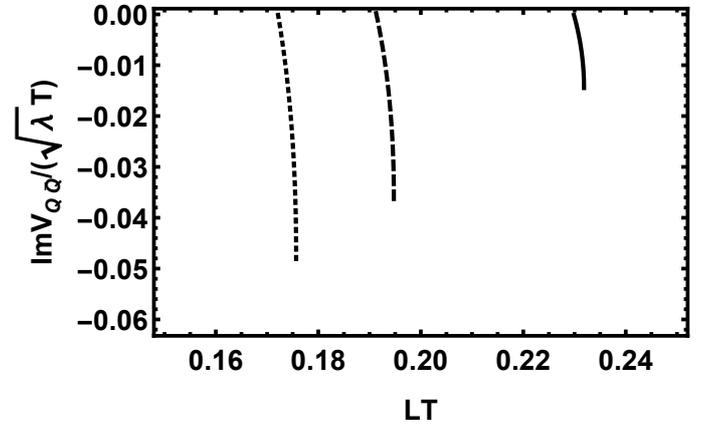}
}
\caption{$ImV/(\sqrt{\lambda}T)$ versus $LT$ for some choices of $c$ with fixed temperature $T = 0.2GeV$. Solid line represents $c = 0.2GeV^4$, dashed line denotes $c = 0.5GeV^4$, dotted line is $c=0.8GeV^4$.}
\label{fig:2}       
\end{figure}
\begin{figure}
\resizebox{0.5\textwidth}{!}{%
  \includegraphics{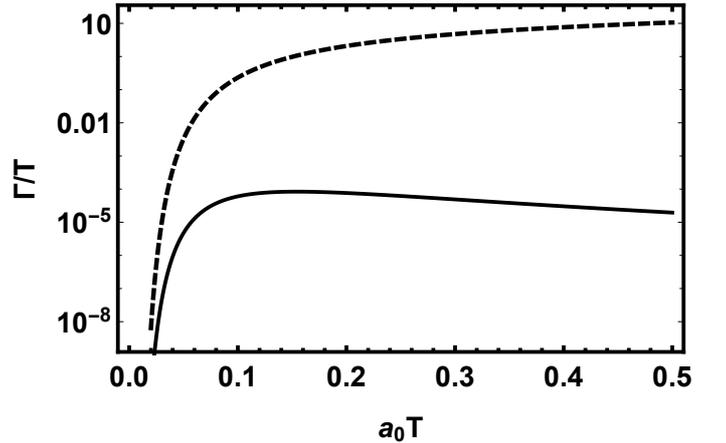}
}
\caption{Compare of exact and approximate approach at $c = 0.2GeV^4$ and T = 0.2GeV. Solid line represents conservative approach, dashed line represents approximate approach.}
\label{fig:3}       
\end{figure}
In Fig.~\ref{fig:2}, we draw $ImV/(\sqrt{\lambda}T)$ against $LT$ for $c=0.2GeV^4,$ $0.5GeV^4$, $0.8GeV^4$. The range of each line of imaginary potential starts at a certain value $LT_{min}$ and ends at $LT_{max}$, which is corresponding to $r_{0max}$ and $r_{0min}$ in Fig.~\ref{fig:1}, respectively. $LT_{min}$ is a point where the imaginary potential becomes negative and $LT_{max}$ is the maximum value during the whole valid range of $rh/r_0$. In addition, we find that increasing gluon condensate causes the imaginary potential to start from smaller distance and have a larger absolute value, which indicates gluon condensate makes the suppression of heavy quarkonium stronger.

In the following, There are two kinds of way to evaluate the thermal width\cite{Fadafan:2013bva,Finazzo:2013aoa,Ali-Akbari:2014vpa}. One approach, named "exact" approach, is only to take the integral range of imaginary potential from $LT_{min}$ to $LT_{max}$. Another approximate approach is to use a straight-line fitting for $ImV$ and the integral range will be from $LT_{min}$ to infinite when evaluating the thermal width. As discussed in these papers, the conservative approach gives a lower bound for the thermal width while the approximate approach will considerablely overestimates the thermal width. It is clearly shown in Fig.~\ref{fig:3}.

 Another point we need to be cautious is the consistence of results calculated in two approaches. Since it is found the non-trivial behavior of thermal width for increasing rapidities, the reason maybe is that the thermal width is dominated by different processes, such as gluo-dissociation and Landau damping for weakly couple plasma\cite{Escobedo:2013tca}. More concretely, the thermal width calculated by conservative approach is decreasing while the thermal width calculated by approximate approach is increasing both for increasing rapidity in holographic calculation\cite{Fadafan:2015kma}.

\begin{figure}
\resizebox{0.5\textwidth}{!}{%
  \includegraphics{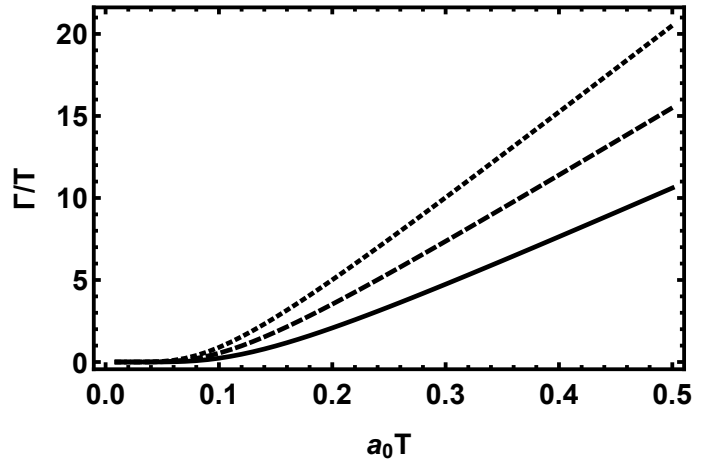}
}
\caption{$\Gamma/T$ as a function of $a_{0}T$ calculated by approximate approach at fixed $T = 0.2GeV$ . Solid line represents $c = 0.2GeV^4$, dashed line denotes $c = 0.5GeV^4$, dotted line is $c=0.8GeV^4$.}
\label{fig:4}       
\end{figure}

\begin{figure}
\resizebox{0.5\textwidth}{!}{%
  \includegraphics{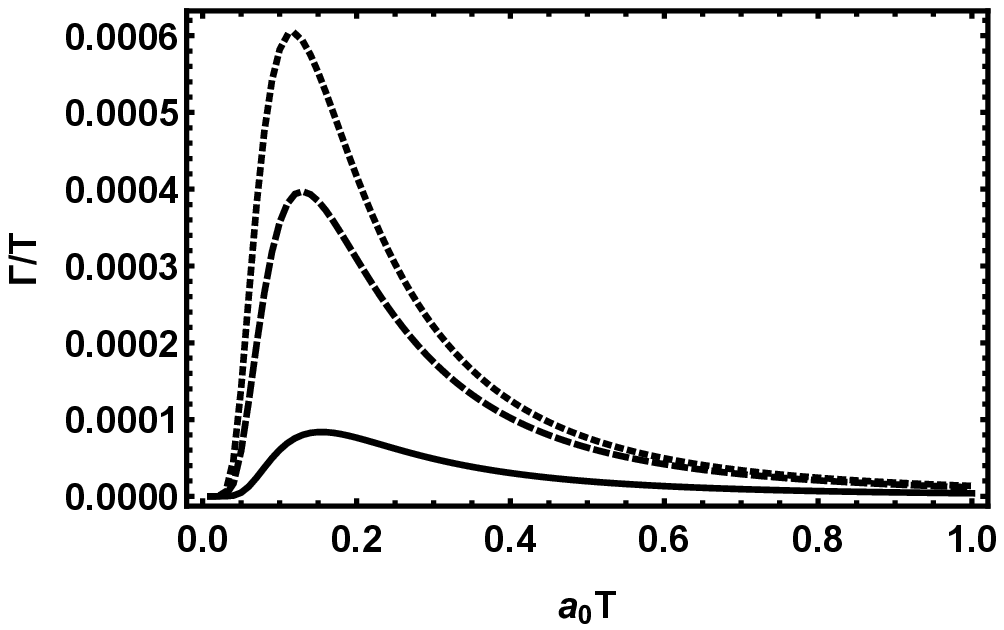}
}
\caption{$\Gamma/T$ as a function of $a_{0}T$ calculated by exact approach at fixed $T = 0.2GeV$ . Solid line represents $c = 0.2GeV^4$, dashed line denotes $c = 0.5GeV^4$, dotted line is $c=0.8GeV^4$.}
\label{fig:5}       
\end{figure}

 For comparison, we use both approaches to evaluate thermal width as shown in Fig.~\ref{fig:4} and Fig.~\ref{fig:5}, respectively. We display $\Gamma/T$ as a function of $a_0T$ for some different values of $c$. Instead of the non-trivial behavior in moving case, it is found that the large value of gluon condensate will lead to large width in both two approaches. Namely, increasing gluon condensate results in a weaker bound for heavy $Q\bar{Q}$ pair. In Ref.\cite{Kim:2008ax}, they investigated the gluon condensate on real potential. they found the heavy quark potential becomes deeper as c decreases, implying dissociation of heavy quarkonium is easy at large gluon condensate. This conclusion is consistent with our analysis of thermal width.

\begin{figure}
\resizebox{0.5\textwidth}{!}{%
  \includegraphics{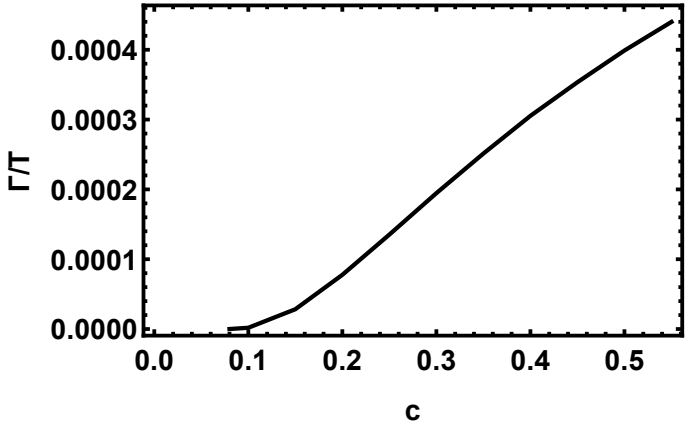}
}
\caption{$\Gamma/T$ of $\Upsilon(1S)$ state as a function of $c$ for fixed $T = 0.2GeV$.}
\label{fig:6}       
\end{figure}

In Fig.~\ref{fig:6}, by setting $m_Q=4.7GeV$, corresponding to $a_0=0.621GeV^{-1}$ and the t'Hooft coupling  $\lambda=9$, we plot the thermal width of $\Upsilon(1S)$ state as a function of the gluon condensate. It is found that there is a critical value $c_{0}$ below which the thermal width will disappear. It can be inferred the critical value of gluon condensate indicating the dissociation of heavy quarkonium. Since, the high temperature leads to a dramatically decreasing of gluon condensate, that is to say, small condensate means very high temperature. But the relationship between temperature and gluon condensate is not related in this metric.

In this work, there are a lot of constraints but the most important constraint is small thermal fluctuations for the saddle point approximation, called as mathematical approximation. In addition, there is a constrain imposed by physics itself, named as physical approximation, existing a threshold $c_{0}$ for fixed temperature, in which the imaginary potential will disappear, equivalently, thermal width will disappear. In some sense, it may indicate the dissociation of heavy quarkonium.


\section{Conclusion and discuss}
\label{sec:Con}
In this paper, we have used the string worldsheet fluctuation approach around the deepest point of the string to study the effect of gluon condensate on imaginary potential of heavy quarkonium in strongly coupled quark gluon plasma by using the AdS/CFT correspondence. The imaginary potential and thermal width are calculated by using a general formula, as shown in~\eqref{eq:x30} and ~\eqref{eq:x33}. A accurate regime of this methods has also been given, $LT_{min}\le\omega\le LT_{max}$, where $LT_{min}$ is determined by $\sigma_c$ which is a real, and $LT_{max}$ corresponds to the maximum of $LT$. The thermal width is calculated by a ground-state wave function of a particle in a Coulomb-like potential and the Bohr radius is inversely proportional with one half mass of ground-state particle.

One can find that the decrease of gluon condensate enlarges the inter-distances and makes the onset of the imaginary potential happen for larger $LT$ which means that the suppression becomes weaker. In addition, the dropping gluon condensate reduces the absolute value of imaginary potential.

Since two different approaches of evaluating thermal width give two opposite behavior in non-vanishing rapidity, for security, we compute the thermal width with these two approaches in non-vanishing gluon condensate and find the consistence of the two approaches. Thus, we can conclude that increasing the value of gluon condensate leads to increasing thermal width. Namely, the dropping gluon condensate makes heavy quarkonium have a stronger bound for fixed tempareture, which shows the heavy quarkonium is easier to dissociation. This conclusion is in agreement with Ref. \cite{Kim:2008ax}, in which they investigate the effect of gluon condensate on real potential. We also find there is a critical value of gluon condensate below which the thermal width will no longer exist. We presume it relates to the dissociation of heavy quarkonium.

The drawback of this model is that the value of gluon condensate don't have a direct connection with temperature. In some sense, the condensation is piecewise constant function, which allows us to consider the effect of gluon condensate and temperature on thermal width separately\cite{Kim:2007qk}. But a real situation requires us to examine the back reaction of gluon condensate and temperature in a more complicated holographic model\cite{DeWolfe:2010he}. This problem can be further discussed in future work.
\section*{acknowledgement}
 We would like to thank Zi-qiang Zhang for useful discussions of imaginary potential.

\bibliographystyle{spphys}
\bibliography{ref}
\end{document}